\documentclass[prd,a4paper,nofootinbib]{revtex4}
\usepackage{epsfig}
\newcommand{\be}{\begin{equation}}
\newcommand{\ee}{\end{equation}}
\newcommand{\bea}{\begin{eqnarray}}
\newcommand{\eea}{\end{eqnarray}}
\newcommand{\bean}{\begin{eqnarray*}}
\newcommand{\eean}{\end{eqnarray*}}

\newcommand{\gapproxeq}{\lower
.7ex\hbox{$\;\stackrel{\textstyle >}{\sim}\;$}}
\newcommand{\lapproxeq}{\lower
.7ex\hbox{$\;\stackrel{\textstyle <}{\sim}\;$}}

\begin{document}

\bibliographystyle{unsrt}

\title{\bf Study of electromagnetic decay of $J/\psi$ and $\psi^\prime$ to vector and pseudoscalar}

\author{Qiang Zhao$^{1,2}$, Gang Li$^1$}
\affiliation{1) Institute of High Energy Physics, Chinese Academy
of Sciences, Beijing, 100049, P.R. China}

\affiliation{2) Department of Physics, University of Surrey,
Guildford, GU2 7XH, United Kingdom}

\author{Chao-Hsi Chang$^3$}

\affiliation{3) Institute of Theoretical Physics, Chinese Academy
of Sciences, Beijing, 100080, P.R. China}

\date{\today}

\begin{abstract}

The electromagnetic decay contributions to $J/\psi(\psi^\prime)\to
VP$, where $V$ and $P$ stand for vector and pseudoscalar meson,
respectively, are investigated in a vector meson dominance (VMD)
model. We show that $J/\psi (\psi^\prime)\to \gamma^*\to VP$ can
be constrained well with the available experimental information.
We find that this process has significant contributions in
$\psi^\prime\to VP$ and may play a key role in understanding the
deviations from the so-called ``12\% rule" for the branching ratio
fractions between $\psi^\prime\to VP$ and $J/\psi\to VP$. We also
address that the ``12\% rule" becomes very empirical in exclusive
hadronic decay channels.

\end{abstract}

\maketitle





\vspace{1cm}


The decay of charmonia into light hadrons is rich of information
about QCD strong interactions between quarks and gluons. Due to
the flavor change in the $c\bar{c}$ annihilation, it is also ideal
for the study of light hadron production mechanisms, and useful
for probing their flavor and gluon contents, such as the search
for experimental evidence for glueball and hybrid. In the past
decade, data for $J/\psi$ decays have experienced a drastic
improvement. We now not only have access to small branching ratio
at order of $10^{-6}$, but also have much precise measurements of
most of those old channels from BES, DM2 and Mark-III. Such a
significant improvement will allow a systematic analysis of
correlated channels, from which we expect that dynamical
information about the light hadron production mechanisms can be
extracted.

In this work, we will study the electromagnetic (EM) decay of
vector charmonia ($J/\psi$ and $\psi^\prime$) into light vector
and pseudoscalar. From an empirical viewpoint, one can separate
the decays of $J/\psi (\psi^\prime)\to VP$ into two classes: i)
Isospin conserved channels such as $J/\psi(\psi^\prime)\to
\rho\pi$, $K^*\bar{K}$, $\omega\eta$, $\phi\eta$, etc. These are
decays via both strong and EM transitions; ii) Isospin violated
channels such as $J/\psi (\psi^\prime) \to \rho\eta$,
$\rho\eta^\prime$, $\omega\pi^0$, and $\phi\pi$, of which the
leading decay amplitudes are from EM transitions. In association
with the above separation is the observation that branching ratios
for some of those isospin violated channels~\cite{pdg2006}, such
as $J/\psi (\psi^\prime) \to \rho\eta$, $\rho\eta^\prime$ and
$\omega\pi^0$, are compatible with the isospin conserved ones such
as $\omega\eta^\prime$ and $\phi\eta^\prime$ in $J/\psi$ decays,
and $\rho\pi$, $\omega\eta$, $\omega\eta^\prime$, $\phi\eta$,
$\phi\eta^\prime$ in $\psi^\prime$ decays. This observation shows
that the EM transition may not be as small as we thought in
comparison with the strong one. Therefore, its roles played in
$J/\psi(\psi^\prime)\to VP$ should be closely investigated.

On a more general ground, the decay channel $J/\psi (\psi^\prime)
\to VP$ has attracted a lot of attention in the literature due to
its property that the characteristic pQCD helicity conservation
rule is violated here~\cite{brodsky-lepage-81}. As a consequence,
the pQCD power suppression occurs in this channel and leads to a
relation for the ratios between $J/\psi$ and $\psi^\prime$
annihilating into three gluons and a single direct photon:
\bea
R &\equiv & \frac{BR(\psi^\prime\to hadrons)}{BR(J/\psi\to
hadrons)}\nonumber\\
&\simeq &\frac{BR(\psi^\prime\to e^+ e^-)}{BR(J/\psi\to e^+
e^-)}\simeq 12\% ,
\eea
which is empirically called ``12\% rule". However, much stronger
suppressions are found in $\rho\pi$ channel, i.e.
$BR(\psi^\prime\to\rho\pi)/BR(J/\psi\to\rho\pi)\simeq (2.0\pm
0.9)\times 10^{-3}$, which gives rise to the so-called ``$\rho\pi$
puzzle". Namely, there exist large deviations from the above
empirical ``12\% rule" in exclusive channels such as $\rho\pi$ and
$K^*\bar{K}+c.c.$

As we know that the ``12\% rule" is based on the expectation that
the charmonium $3g$ strong decays are the dominant ones in
exclusive decay channels, the large deviations from the ``12\%
rule" in $\rho\pi$ and $K^*\bar{K}+c.c.$ naturally imply some
underlying mechanisms which can interfere with the $3g$ decays and
change the branching ratio fractions. Theoretical explanations for
the ``12\% rule" deviations have been proposed in the
literature~\cite{hou-soni,karl-roberts,pinsky-85,brodsky-lepage-tuan,ct,pinsky-90,brodsky-karliner,li-bugg-zou,chen-braaten,gerald,feldman,suzuki,rosner,wym,liu-zeng-li},
but so far none of those solutions has been indisputably
agreed~\cite{yuan,m-y-w}. This makes it necessary to provide a
detailed study of the charmonium EM decays. If compatible strength
of the EM transition occurs in some of those exclusive decay
channels, one can imagine that large interferences between the EM
and strong transitions are possible, and they may be one of the
important sources which produce large deviations from the ``12\%
rule" in $VP$ decay channels. Relevant studies can also be found
in the literature for understanding the role played by the EM
transitions in $J/\psi$ decays. Parametrization schemes are
proposed to estimate the EM decay contributions to $J/\psi\to VP$
in Refs.~\cite{haber,seiden}, but a coherent study of $J/\psi$ and
$\psi^\prime$ is still unavailable.

Apart from the above interests, the EM decay of $J/\psi\to VP$ is
also rich of dynamical information about the Okubo-Zweig-Iizuka
(OZI) rule~\cite{ozi}. The decay of $J/\psi\to \phi\pi^0$ involves
both isospin and OZI-rule violations. Although only the upper
limit, $BR^{exp}(J/\psi\to\phi\pi^0)< 6.4\times 10^{-6}$, is
given, this will be an interesting place to test dynamical
prescriptions for $J/\psi\to \gamma^*\to VP$. In
Refs.~\cite{pinsky-85,pinsky-90}, apart from the EM process, the
OZI doubly disconnected processes are also investigated, which
however possesses large uncertainties. In particular, the
separation of these two correlated processes is strongly
model-dependent.

In this work, we will introduce an effective Lagrangian for $V\gamma
P$ couplings, and apply the vector meson dominance (VMD) model to
$V\gamma^*$ couplings. By studying the $J/\psi(\psi^\prime)\to
\gamma^*\to VP$ at tree level, we shall examine the ``12\% rule" for
those exclusive $VP$ decay channels. Deviations from this empirical
rule in the exclusive decays can thus be highlighted.

\begin{figure}
\begin{center}
\epsfig{file=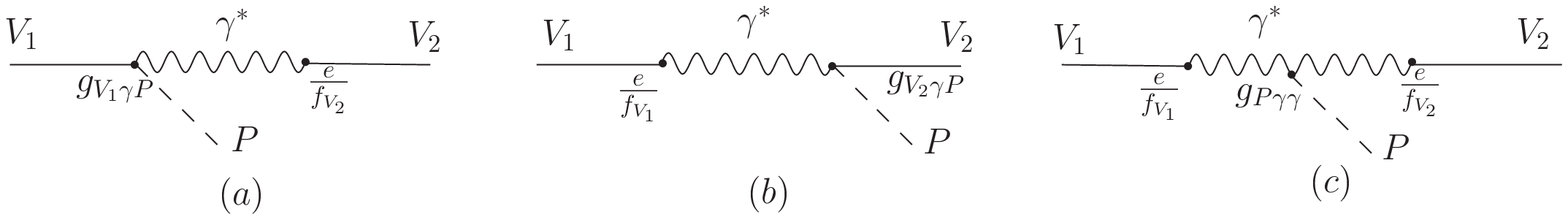, width=12cm,height=2.8cm} \caption{Schematic
diagrams for $J/\psi (\psi^\prime)\to \gamma^*\to VP$. }
\protect\label{fig-1}
\end{center}
\end{figure}

For $J/\psi(\psi^\prime)\to \omega \eta$, $\omega\eta^\prime$,
$\phi\eta$, $\phi\eta^\prime$, $\rho\pi$ and $K^*\bar{K}+c.c.$,
the strong and EM decay process are mixed, and the former
generally plays a dominant role. For $J/\psi (\psi^\prime) \to
\rho\eta$, $\rho\eta^\prime$, $\omega\pi$, and $\phi\pi$, the
transitions are via EM processes, of which the isospin
conservation is violated. Typical transitions for $V_1\to V_2 P$
are illustrated by Fig.~\ref{fig-1}, which consists of three
contributions: (a) the process that the pseudoscalar meson is
produced in association with the virtual photon via $V_1$
annihilation; (b) the pseudoscalar produced at the final state
vector meson $V_2$ vertex; and (c) the pseudoscalar produced via
the axial current anomaly. Note that isospin conservation can be
violated in both Fig.~\ref{fig-1}(a) and (b), with the observation
of non-zero branching ratios for $J/\psi\to \gamma
\pi^0$~\cite{pdg2006}. At hadronic level, these are independent
processes where all the vertices can be determined by other
experimental measurements. This treatment is different from that
of Refs.~\cite{pinsky-85,pinsky-90}. Although the OZI disconnected
diagram considered in Refs.~\cite{pinsky-85,pinsky-90} is similar
to our Fig.~\ref{fig-1}(b), our consideration of the $V\gamma^* P$
coupling will allow us to include both OZI and isospin violation
effects which can be constrained by experimental data.

We introduce a typical effective Lagrangian for the $V\gamma P$
coupling:
\be\label{lagrangian-1}
{\cal L}_{V\gamma P}=\frac{g_{V\gamma
P}}{M_V}\epsilon_{\mu\nu\alpha\beta}\partial^\mu
V^\nu\partial^\alpha A^\beta P
\ee
where $V^\nu(=\rho, \ \omega, \ \phi, \ J/\psi, \
\psi^\prime\dots)$ and $A^\beta$ are the vector meson and EM
field, respectively; $M_V$ is the vector meson mass;
$\epsilon_{\mu\nu\alpha\beta}$ is the anti-symmetric Levi-Civita
tensor.

The $V\gamma^*$ coupling is described by the VMD model,
\be
{\cal L}_{V\gamma}=\sum_V \frac{e M_V^2}{f_V} V_\mu A^\mu \ ,
\ee
where $eM_V^2/f_V$ is a direct photon-vector-meson coupling in
Feynman diagram language, and the isospin 1 and 0 component of the
EM field are both included.

The invariant transition amplitude for $V_1\to\gamma^*\to  V_2 P$
can thus be expressed as: \bea \label{4}
{\cal M} & \equiv & {\cal M}_A + {\cal M}_B + {\cal M}_C \nonumber\\
&=& \left( \frac{e}{f_{V2}}\frac{g_{V1\gamma P}}{M_{V1}}{\cal F}_a
+ \frac{e}{f_{V1}}\frac{g_{V2\gamma P}}{M_{V2}}{\cal F}_b +
\frac{e^2}{f_{V1} f_{V2}} \frac{g_{P\gamma\gamma}}{M_P}{\cal F}_c
\right)\epsilon_{\mu\nu\alpha\beta}\partial^\mu
V_1^\nu\partial^\alpha V_2^\beta P
\eea
where $g_{P\gamma\gamma}$ is the coupling for the neutral
pseudoscalar meson decay to two photons; ${\cal F}_a$ and ${\cal
F}_b$ denote the form factor corrections to the $V\gamma^* P $
vertices in comparison with the real photon transition for $V\to
\gamma P$; and ${\cal F}_c$ is the form factor for $P\to
\gamma^*\gamma^*$.

The partial decay width can thus be expressed as \bea \label{5}
\Gamma(V_1\to V_2P) &=&\frac{|{\bf p}_{v2}|}{8\pi
M_{V1}^2}\frac{1}{2S+1}\sum|{\cal M}|^2 \nonumber\\
&=&\frac{|{\bf p}_{v2}|^3}{12\pi}\left(
\frac{e}{f_{V2}}\frac{g_{V1\gamma P}}{M_{V1}}{\cal F}_a +
\frac{e}{f_{V1}}\frac{g_{V2\gamma P}}{M_{V2}}{\cal F}_b +
\frac{e^2}{f_{V1} f_{V2}} \frac{g_{P\gamma\gamma}}{M_P}{\cal
F}_c\right)^2 \ ,
\eea
where $S=1$ is the spin of the initial vector meson; $|{\bf
p}_{v2}|$ is the three-momentum of the final state vector meson in
the initial-vector-meson-rest frame.

Those three typical coupling constants are determined as follows:

(I) For $V\to \gamma P$ decay, following the effective Lagrangian
of Eq.~(\ref{lagrangian-1}), we derive the coupling constant:
\be
g_{V\gamma P}=\left[\frac{12\pi M_V^2\Gamma^{exp}(V\to \gamma
P)}{|{\bf p}_\gamma|^3}\right]^{1/2} ,
\ee
where $|{\bf p}_\gamma|$ is the three-momentum of the photon in
the initial vector meson rest frame; $\Gamma^{exp}(V\to \gamma P)$
is the vector meson radiative decay partial width, and available
in experiment.

For $\rho\gamma\eta^\prime$ and $\omega\gamma \eta^\prime$
couplings, we determine the coupling constants in $\eta^\prime \to
\gamma \rho$ and $\gamma \omega$:
\be
g_{V\gamma \eta^\prime}=\left[\frac{4\pi
M_V^2\Gamma^{exp}(\eta^\prime\to \gamma V)}{|{\bf
p}_\gamma^\prime|^3}\right]^{1/2} ,
\ee
where $|{\bf p}_\gamma^\prime|$ is the three-momentum of the
photon in the $\eta^\prime$ rest frame.

(II) The $V\gamma^*$ coupling is determined in $V\to e^+ e^-$
channel. With the partial decay width $\Gamma_{V\to e^+ e^-}$, the
coupling constant $e/f_V$ can be derived:
\be
\frac{e}{f_V}=\left[\frac{3\Gamma_{V\to e^+ e^-}}{4\alpha_e |{\bf
p}_e|}\right]^{1/2} ,
\ee
where we have neglected the mass of the electron and $|{\bf p}_e|$
is the electron three-momentum in the vector meson rest frame;
$\alpha_e=1/137$ is the fine-structure constant.

(III) For $P\to \gamma\gamma$, we adopt the following form of
effective Lagrangian: \be {\cal
L}_{P\gamma\gamma}=\frac{g_{P\gamma\gamma}}{M_P}
\epsilon_{\mu\nu\alpha\beta}\partial^\mu A^\nu\partial^\alpha
A^\beta P \ , \ee where the coupling constant is normalized to the
pseudoscalar meson mass $M_P$. With the partial decay width
$\Gamma^{exp}(P\to\gamma\gamma)$ the coupling constant for real
photon in the final state can be derived: \be
g_{P\gamma\gamma}=[32\pi \Gamma^{exp}(P\to \gamma\gamma)/M_P]^{1/2}
\ . \ee

It is encouraging that for all the decay channels of $J/\psi
(\psi^\prime) \to \gamma^*\to VP$, the experimental data are
available for determining the above coupling constants:
$g_{V\gamma P}$, $e/f_V$, and $g_{P\gamma\gamma}$~\cite{pdg2006}.
We are then left with the only uncertainty from the form factors
due to the exchange of off-shell photons.

We find that without form factors, i.e. ${\cal F}_a={\cal
F}_b={\cal F}_c=1$, the calculated branching ratios for the
isospin violated channels will be significantly overestimated.
This is expected due to the large virtualities of the off-shell
photons and the consequent power suppressions from the pQCD
hadronic helicity-conservation~\cite{brodsky-lepage-81}. Since we
think that the non-perturbative QCD effects might have played a
role in the transitions at $J/\psi$ energy\footnote{For instance
such as in the inclusive decay $J/\psi\to q\bar{q}$, the
`virtualness' of one quark in the created quark pair is less than
chiral broken energy scale $\Lambda_\chi\sim 1.0$ GeV, thus, the
non-perturbative QCD effects must be sizeable in the concerned
decays here.}, e.g. in Fig.~\ref{fig-1}(a) and (b) a pair of
quarks may be created from vacuum as described by $^3P_0$ model,
the pQCD hadronic helicity-conservation due to the vector nature
of gluon is violated quite strongly, thus alternatively, we would
like to suggest a monopole-like (MP) form factor dedicated to the
suppression effects:
\be\label{mp}
{\cal F}(q^2)=\frac{1}{1-q^2/\Lambda^2} \ ,
\ee
where $\Lambda$ can be regarded as an effective mass accounting
for the overall effects from possible resonance poles and
scattering terms in the time-like kinematic region, and will be
determined by fitting the data~\cite{pdg2006} for $J/\psi
(\psi^\prime) \to \rho\eta$, $\rho\eta^\prime$, $\omega\pi^0$, and
$\phi\pi^0$.

It should be noted that this MP form factor can only partly depict
the pQCD power suppression due to violations of the hadronic
helicity conservation when $q^2\gg
\Lambda^2$~\cite{brodsky-lepage-81}, but it is quite consistent
with the VMD framework.

By adopting the MP empirical form factor, we have already assumed
that non-perturbative effects might have played a substantial role
in the transitions.  In principle it should be tested
experimentally via measuring the coupling the processes
$J/\psi(\psi^\prime)\to P e^+ e^-$ and $e^+ e^-\to P e^+ e^-$,
respectively, when the integrated luminosity at $J/\psi$ and the
suitable energies for $e^+e^-$ colliders is accumulated enough.

The form factor ${\cal F}_c$ appearing in Eqs.~(\ref{4}) and
(\ref{5}) can be determined in $\gamma^*\gamma^*$ scatterings. A
commonly adopted form factor is \be {\cal
F}_c(q_1^2,q_2^2)=\frac{1}{(1-q_1^2/\Lambda^2)(1-q_2^2/\Lambda^2)}
\ , \ee where $q_1^2=M_{V1}^2$ and $q_2^2=M_{V2}^2$ are the
squared four-momenta carried by the time-like photons. We assume
that the $\Lambda$ is the same as in Eq.~(\ref{mp}), thus, ${\cal
F}_c={\cal F}_a {\cal F}_b$.

Proceeding to the numerical calculations, we first determine the
coupling constants, $e/f_V$, $g_{V\gamma P}$ and
$g_{P\gamma\gamma}$ in the corresponding decays, and the results
are listed in Tables~\ref{tab-1}-\ref{tab-3}. It shows that the
$e/f_\rho$ coupling is the largest one while all the others are
compatible. For the $g_{V\gamma P}$, it is sizeable for light
vector mesons and much smaller for $J/\psi$ and $\psi^\prime$.
Note that there is no datum for $\psi^\prime\to \gamma \pi$
available. So we simply put $g_{\psi^\prime\gamma\pi}=0$ in the
calculations. The $P\gamma\gamma$ couplings can be well determined
due to the good shape of the data~\cite{pdg2006}.

To examine the role played by the form factors, we first calculate
the EM decay branching ratios without form factors, i.e. ${\cal
F}(q^2)=1$. It shows that all the data are significantly
overestimated by the theoretical predictions as shown by
Tables~\ref{tab-4} and \ref{tab-5}. Nonetheless, it shows that
without form factors process (b) is the only dominant transition.

To determine the effective mass $\Lambda$ in the MP form factor,
we consider two possible relative phases between process (a) and
(b) in fitting the data for the isospin violated channels, $J/\psi
(\psi^\prime)\to \rho\eta$, $\rho\eta^\prime$, and $\omega\pi$. We
mention in advance that the contributions from process (c) will
bring only few percent corrections to the results. Since the
corrections are within the datum uncertainties, we are not
bothered to consider its relative phase to process (a) and (b). In
Tables~\ref{tab-4} and \ref{tab-5}, the results for process (a)
and (b) in a constructive phase (MP-C) with $\Lambda=0.616\pm
0.008$ GeV, and in a destructive phase (MP-D) with
$\Lambda=0.65\pm 0.01$ GeV are listed. The reduced $\chi^2$ values
are $\chi^2=4.1$ in MP-C and 14.2 in MP-D, respectively.

With the above fitted values for $\Lambda$ (MP-C and MP-D),
predictions for those isospin conserved channels in $J/\psi\to
\gamma^*\to VP$ and $\psi^\prime\to\gamma^*\to VP$ are listed in
Table~\ref{tab-4} and \ref{tab-5} to compare with the experimental
data~\cite{pdg2006}.

There arise some basic issues from the theoretical results.

(I) We find that even though with the form factors, process (b) in
Fig.~\ref{fig-1} is still the dominant one in most channels except
for $\rho\eta^\prime$. For most channels the couplings
$g_{J/\psi\gamma P}$ and $g_{\psi^\prime\gamma P}$ in process (a)
are generally small, and similarly are $e^2/(f_{V1}f_{V2})$ and
$g_{P\gamma\gamma}$ in process (c). However, we find that the
amplitudes of process (a) and (b) are compatible in $J/\psi\to
\rho\eta^\prime$. As shown in Table~\ref{tab-4}, large
cancellations appear in the branching ratio when (a) and (b) are
in a destructive phase (Column MP-D). This is due to the
relatively large branching ratios for
$J/\psi\to\gamma\eta^\prime$~\cite{pdg2006}. Such a large
difference between these two phases makes the $\rho\eta^\prime$
channel extremely interesting. The branching ratio fraction will
be useful for distinguishing the relative phases between (a) and
(b) in the isospin violated channels. It also highlights the
empirical aspect of the pQCD ``12\% rule" in exclusive hadronic
decays.

We also note that process (a) and (c) do not contribute to
$K^*\bar{K}+c.c.$ and $\rho^+\pi^- +c.c.$ This turns to be an
advantage for understanding the decay mechanism of
$J/\psi(\psi^\prime)\to \gamma^*\to \rho^+\pi^- +c.c.$ and
$K^*\bar{K}+c.c.$, and should be also an ideal place to test the
``12\% rule" in exclusive decays.

To be more specific, we analyze first those four isospin violation
decays: $J/\psi (\psi^\prime)\to \gamma^*\to \rho\eta$,
$\rho\eta^\prime$, $\omega\pi^0$ and $\phi\pi^0$. These decays to
leading order are through EM transitions. Transitions of
Fig.~\ref{fig-1} have shown how the kinematic and form-factor
corrections can correlate with the naive pQCD expected ratio:
$\Gamma(\psi^\prime\to e^+ e^-)/\Gamma(J/\psi\to e^+ e^-)$, i.e.
the ``12\% rule", and makes it very empirical.

As an example, for those channels dominated by process (b), the
exclusive decays are still approximately proportional to the
charmonium wavefunction at its origin, i.e. $|\psi(0)|^2$, by
neglecting the contributions from process (a) and (c). The
branching ratio fraction can be expressed as:
\bea
\label{rvp} R^{VP} &\equiv &\frac{BR(\psi^\prime\to\gamma^*\to
VP)}{BR(J/\psi\to\gamma^*\to VP)} \nonumber\\
&\simeq &\frac{BR(\psi^\prime \to e^+ e^-)}{BR(J/\psi\to e^+
e^-)}\frac{|{\bf p}_e|}{|{\bf p}_e^\prime|}\frac{|{\bf
p}_{v2}^\prime|^3}{|{\bf p}_{v2}|^3}\frac{{\cal
F}_b^2(M_{\psi^\prime}^2)}{{\cal F}_b^2(M_{J/\psi}^2)},
\eea
where $|{\bf p}_e|$ and $|{\bf p}_e^\prime|$ are three-momenta of
the electron in $J/\psi\to e^+ e^-$ and $\psi^\prime\to e^+ e^-$
in the vector meson rest frame, respectively; while $|{\bf
p}_{v2}|$ and $|{\bf p}_{v2}^\prime|$ are three momenta of the
final state vector mesons in $J/\psi\to V_2 P$ and $\psi^\prime
\to V_2 P$, respectively. It shows that the respect of the ``12\%
rule" requires that the kinematic and form factor corrections
cancel each other for all those channels, which however, is not a
necessary consequence of the physics at all. Including the
contributions from process (a) and (c) will worsen the situation.

To see this more clearly, we list the branching ratio fractions
for the choice of MP-C ($R_3^{VP}$), MP-D ($R_2^{VP}$) and {\it
without} form factors ($R_1^{VP}$) in Table~\ref{tab-6} to compare
with the data. It shows that {\it without} the form factor
corrections, ratio $R_1^{VP}$ has values in a range of $(19\sim
21)\%$ for those four channels, which are larger than the
expectation of the ``12\% rule".

With the form factors, it shows that $R_3^{VP}$ has a stable range
of $(7\sim 9)\%$, while more drastic changes occur to $R_2^{VP}$.
For instance, we obtain $R_2^{\rho\pi}=52\%$ which strongly
violates the ``12\% rule". Unfortunately, the data still have
large uncertainties. We expect that an improved branching ratio
fraction for this channel will be able to determine the relative
phase between process (a) and (b) in our model, and highlight the
underlying mechanism.

The branching ratios for $\phi\pi^0$ channel are much smaller than
others due to the small $\phi\gamma\pi$ coupling. This is in a
good agreement with the OZI rule suppressions expected in
$\phi\pi^0$ channel.

(III) For the isospin conserved channels, the EM decay
contributions in $J/\psi$ decays turn out to be rather small in
both MP-D and MP-C phases. This is consistent with studies in the
literature that $J/\psi\to VP$ is dominated by the $3g$
transitions. Thus, the deviation of the ``12\% rule" could be more
likely due to the suppression of the amplitudes in $\psi^\prime\to
VP$ (see the review of Ref.~\cite{yuan}).

If we simply apply the relations between the strong and EM
transitions parametrized by Ref.~\cite{seiden}, the ratio between
charged and neutral channels can be expressed as:
\be
Q\equiv \frac{BR(\psi^\prime\to K^{*+}K^-+c.c.)}{BR(\psi^\prime\to
K^{*0}\bar{K^0}+c.c.)}\simeq \frac{[g(1-s)+e]^2}{[g(1-s)-2e]^2} \
,
\ee
where $g$ and $e$ denote the strong and EM decay strengths,
respectively, and $s\simeq 0.1$ is a parameter for the flavor
SU(3) breaking. One can see that for $e = (-1/3 \sim -1/2)\times
g(1-s)$, we have $Q\simeq 0.06\sim 0.16$, which is in a good
agreement with the data, $0.08\sim 0.28$~\cite{pdg2006}.

We can also check the other two correlated relations:
\be
\frac{BR(\psi^\prime\to \gamma^*\to
K^{*+}K^-+c.c.)}{BR(\psi^\prime\to K^{*+}K^-+c.c.)}\simeq
\frac{e^2}{[g(1-s)+e]^2}\simeq 0.25\sim 1,
\ee
corresponding to $e = (-1/3 \sim -1/2)\times g(1-s)$. This is
consistent with the range of $BR^{MP}_C/BR^{exp}\simeq 0.22\sim
0.56$ and $BR^{MP}_D/BR^{exp}\simeq 0.28\sim 0.70$.

Similarly, for $\psi^\prime\to K^{*0}\bar{K^0}+c.c.$ we have
\be
\frac{BR(\psi^\prime\to \gamma^*\to
K^{*0}\bar{K^0}+c.c.)}{BR(\psi^\prime\to
K^{*0}\bar{K^0}+c.c.)}\simeq \frac{4e^2}{[g(1-s)-2e]^2}\simeq
0.16\sim 0.44 ,
\ee
corresponding to the same range for $e$ ($e = (-1/3 \sim
-1/2)\times g(1-s)$). It also turns to be compatible with
$BR^{MP}_C/BR^{exp}\simeq 0.10 \sim 0.15$ and
$BR^{MP}_D/BR^{exp}\simeq 0.12 \sim 0.18$.

For $\rho\pi$ channel, the above relative phase between the strong
and EM transitions can explain the relatively small branching
ratios for $\psi^\prime\to \rho\pi$, i.e. the EM amplitude will
destructively interfere will the strong one. With $e=(-1/3\sim
-1/2)g(1-s)\simeq (-1/3\sim -1/2)g$~\cite{seiden}, we have a
relation:
\be
\frac{BR(\psi^\prime\to\gamma^*\to\rho\pi)}{BR(\psi^\prime\to
\rho\pi)} \simeq \frac{e^2}{(g+e)^2}=0.25\sim 1 ,
\ee
which is also consistent with $BR^{MP}_C/BR^{exp}=0.18\sim 0.39$
and $BR^{MP}_D/BR^{exp}=0.22\sim 0.49$~\cite{pdg2006}.

The above analysis suggests that the relative phase between the EM
and strong transition amplitudes in $\psi^\prime\to \rho\pi$
favors $180^\circ$. But due to the large uncertainties with the
data, other phases may be possible~\cite{wym-2004}.

Evidently, for those channels of which the branching ratio
fractions between $\psi^\prime$ and $J/\psi$ are observed to
deviate from the ``12\% rule" (see Table~\ref{tab-6}), such as
$\rho\pi$ and $K^*\bar{K}+c.c.$, they turn to have sizeable EM
contributions as shown by Tables~\ref{tab-4} and \ref{tab-5}.
Since the $J/\psi$ decays are still dominated by the strong
transitions, the measured branching ratio fractions mostly reflect
the interfered effects in $\psi^\prime$ decays over the strong
transitions in $J/\psi\to VP$.

We also analyze EM decay contributions to $\psi(3770)\to VP$, and
find they becomes negligibly small due to the small partial width
for $\psi(3770)\to e^+ e^-$ and form factor suppressions. In most
channels, the EM transitions contribute to the branching ratios at
$O(10^{-8})$ or even smaller, which is beyond the access of the
present experimental facilities. Therefore, we can conclude that
for those measured $\psi(3770)\to VP$ channels the isospin
violations must be negligibly small.

This approach can also be applied to the study of the EM decay
contributions to $\phi\to \rho\pi$ and $\omega\pi$. For the
isospin violated $\omega\pi^0$ channel, with the same form factor
(MP-C), we find $BR(\phi\to\omega\pi^0)=1.8\times 10^{-6}$ and
$BR(\phi\to\rho\pi)=3.2\times 10^{-6}$ in contrast with the the
experimental data,
$BR^{exp}(\phi\to\omega\pi^0)=\left(5.2\begin{array}{c}+1.3\\ -1.1
\end{array}\right) \times 10^{-5}$ and $BR^{exp}(\phi\to\rho\pi+\pi^+\pi^-\pi^0)=(15.3\pm 0.4)\%$~\cite{pdg2006}.
Note that $\phi$ and $\omega$ are quite close in mass, the results
should be sensitive to the form factors, and more elaborate
treatment for them are generally required. If we slightly adjust
$\Lambda$, we can reproduce the data for $\omega\pi^0$ well. For
$\rho\pi$ channel the EM contributions are found much smaller than
the data which suggests the dominance of strong decays in
$\phi\to\rho\pi$.

In summary, we make a systematic analysis of the EM decay
contributions to $J/\psi (\psi^\prime)\to \gamma^*\to VP$
essentially in a VMD model. With the constraint from the available
independent experimental data, the EM contributions in all the
$VP$ hadronic channels can be consistently investigated. Our
results are in a good agreement with the data for those isospin
violated channels such as $\rho\eta$, $\rho\eta^\prime$,
$\omega\pi$, and $\phi\pi$, especially for the `MP-C' model (see
Tables~\ref{tab-4},\ref{tab-5} and \ref{tab-6}). For most of those
isospin conserved channels, we find that the EM decay
contributions can bring sizeable corrections to the
$\psi^\prime\to VP$, which could be a major source accounting for
the branching ratio fraction deviations between the $VP$ exclusive
decays of $J/\psi$ and $\psi^\prime$. Large difference between the
data for $\psi^\prime\to K^{*0}\bar{K^0}+c.c.$ and $K^{*+}K^-
+c.c.$~\cite{pdg2006} has shown such a possibility. An improved
measurement of these two channels will be able to clarify the
predominant role played by the EM transitions.

For the $\rho\pi$ channel, we find that the EM decay contributions
in $\psi^\prime$ decays is compatible with the experimental data.
This highlights that the interferences from the EM transitions can
play essential role in the decay transitions, especially, in the
understanding of the abnormally small ratio of $BR(\psi^\prime\to
\rho\pi)/BR(J/\psi\to \rho\pi)$. However, it should be pointed out
that the compatible strength of the EM decay contributions with the
experimental data in $\psi^\prime\to \rho\pi$ may reflect the
suppressed strong decay strength. In this sense, a full
understanding of the small branching ratio fraction in $\rho\pi$
channel will require a coherent study of the strong decay
mechanisms~\cite{li-bugg-zou,rosner,wym}. This is beyond the focus
of this work and should be pursued in further studies.

Finally, we note that a full constraint on this model can be
reached by accommodating experimental measurements of
$J/\psi(\psi^\prime)\to P e^+ e^-$ and $e^+ e^-\to P e^+ e^-$ to
derive the couplings for $g_{V_1\gamma P}$ and $g_{V_2\gamma P}$
from which the form factor information can be derived. Ambiguities
from the empirical form factors can thus be minimized.
Experimental analysis of these processes at BES is thus strongly
recommended.

\section*{Acknowledgement}

The authors thank C.Z. Yuan, B.S. Zou and H.B. Li for useful
discussions. Q.Z. acknowledges support from the U.K. EPSRC (Grant
No. GR/S99433/01) and the Institute of High Energy Physics of
Chinese Academy of Sciences. C.H.C is supported by National
Natural Science Foundation of China (Grant No.10547001 and
No.90403031).

\begin{table}[ht]
\begin{tabular}{c|c|c|c}
\hline Coupling const. $e/f_V$ & Values ($\times 10^{-2}$) & Total
width of $V$
& $BR(V\to e^+ e^-)$  \\[1ex]
\hline $e/f_\rho$ & 4.28 & 146.4 MeV & $(4.70\pm
0.08)\times 10^{-5}$ \\[1ex]
$e/f_\omega$ & 1.26 & 8.49 MeV & $(7.18\pm
0.12)\times 10^{-5}$\\[1ex]
$e/f_\phi$ & 1.60  & 4.26 MeV & $(2.97\pm
0.04)\times 10^{-4}$ \\[1ex]
$e/f_{J/\psi}$ & 1.92 & $93.4$ keV & $(5.94\pm 0.06)\%$  \\[1ex]
$e/f_{\psi^\prime}$ & 1.17 & 337 keV & $(7.35\pm 0.18)\times
10^{-3}$ \\[1ex] \hline
\end{tabular}
\caption{ The coupling constant $e/f_V$ determined in $V\to e^+
e^-$.  The data for branching ratios are from
PDG2006~\cite{pdg2006}. } \label{tab-1}
\end{table}

\begin{table}[ht]
\begin{tabular}{c|c|c}
\hline Coupling const. $g_{V\gamma P}$ & Values
& Branching ratios  \\[1ex]\hline
$g_{\rho\gamma \eta}$ & 0.372 & $(2.95\pm
0.30)\times 10^{-4}$ \\[1ex]
$g_{\rho\gamma \eta^\prime}$ & 0.302 & $(29.4\pm 0.9)\%$ \\[1ex]
$g_{\rho^0\gamma \pi^0}$ & 0.197 & $(6.0\pm 0.8)\times 10^{-4}$
\\[1ex]
$g_{\rho^\pm\gamma \pi^\pm}$ & 0.170 & $(4.5\pm 0.5)\times
10^{-4}$
\\[1ex]
$g_{\omega\gamma \eta}$ & 0.110 & $(4.9\pm 0.5)\times 10^{-4}$ \\[1ex]
$g_{\omega\gamma\eta^\prime}$ & 0.107 & $(3.03\pm 0.31)\%$ \\[1ex]
$g_{\omega\gamma\pi}$ & 0.565 & $(8.90\begin{array}{c} +0.27
\\-0.23 \end{array})\%$ \\[1ex]
$g_{\phi\gamma\eta}$ & 0.213 & $(1.301\pm 0.024)\%$ \\[1ex]
$g_{\phi\gamma\eta^\prime}$ & 0.218 & $(6.2\pm 0.7)\times 10^{-5}$
\\[1ex]
$g_{\phi\gamma\pi}$ & 0.041 & $(1.25\pm 0.07)\times 10^{-3}$
\\[1ex]
$g_{K^{*\pm}\gamma K^\pm}$ & 0.225 & $(9.9\pm 0.9)\times 10^{-4}$ \\[1ex]
$g_{K^{*0}\gamma K^0}$ & 0.340 & $(2.31\pm 0.20)\times 10^{-3}$
\\[1ex]
$g_{J/\psi\gamma\eta}$ & $3.13\times 10^{-3}$ & $(9.8\pm
1.0)\times 10^{-4}$ \\[1ex]
$g_{J/\psi\gamma\eta^\prime}$ & $7.61\times 10^{-3}$ & $(4.71\pm
0.27)\times 10^{-3}$ \\[1ex]
$g_{J/\psi\gamma\pi}$ & $5.49\times 10^{-4}$ &
$(3.3\begin{array}{c} +0.6 \\ -0.4 \end{array} )\times 10^{-5}$
\\[1ex]
$g_{\psi^\prime\gamma\eta}$ & $1.63\times 10^{-3}$ & $< 9\times
10^{-5}$ \\[1ex]
$g_{\psi^\prime\gamma\eta^\prime}$ & $2.26\times 10^{-3}$ &
$(1.5\pm 0.4)\times 10^{-4}$ \\[1ex]
 \hline
\end{tabular}
\caption{ The coupling constant $g_{V\gamma P}$ determined in
$V\to \gamma P$ or $P\to \gamma V$. The data for branching ratios
are from PDG2006~\cite{pdg2006}. } \label{tab-2}
\end{table}

\begin{table}[ht]
\begin{tabular}{c|c|c}
\hline Coupling const. $g_{P\gamma\gamma}$ & Values
& Branching ratios  \\[1ex]\hline
$g_{\pi\gamma\gamma}$ & $2.40\times 10^{-3}$ & $(98.798\pm 0.032)\%$ \\[1ex]
$g_{\eta\gamma\gamma}$ & $9.70\times 10^{-3}$ & $(39.38\pm 0.26)\%$ \\[1ex]
$g_{\eta^\prime\gamma\gamma}$ & $2.12\times 10^{-2}$ & $(2.12\pm 0.14)\% $ \\[1ex]
 \hline
\end{tabular}
\caption{ The coupling constant $g_{P\gamma\gamma}$ determined in
$P\to \gamma\gamma$. The data for branching ratios are from
PDG2006~\cite{pdg2006}. } \label{tab-3}
\end{table}

\begin{table}[ht]
\begin{tabular}{c|c|c|c|c}
\hline Decay channels & ${\cal F}(q^2)=1$ & MP-D& MP-C   & Exp.
data  \\[1ex]\hline
$\rho\eta$ & $6.8\times 10^{-2}$ & $8.0\times 10^{-5}$ &
$1.6\times 10^{-4}$ &
$(1.93\pm 0.23)\times 10^{-4}$ \\[1ex]
$\rho\eta^\prime$ & $3.5\times 10^{-2}$ & $4.4\times 10^{-6}$  &
$1.5\times 10^{-4}$ &
$(1.05\pm 0.18)\times 10^{-4}$ \\[1ex]
$\omega\pi$ & $0.16$ & $3.4\times 10^{-4}$ & $2.8\times 10^{-4}$ &
$(4.5\pm 0.5)\times
10^{-4}$ \\[1ex]
$\phi\pi$ & $4.4\times 10^{-4}$ &  $8.1\times 10^{-7}$ &
$8.3\times 10^{-7}$  & $<
6.4\times 10^{-6}$ \\[1ex]\hline
$\rho^0\pi^0$ & $2.0\times 10^{-2}$ & $3.6\times 10^{-5}$ &
$3.9\times 10^{-5}$ &
$(5.6\pm 0.7)\times 10^{-3}$ \\[1ex]
$\rho\pi$ & $ 5.2\times 10^{-2}$ & $1.0\times 10^{-5}$  &
$9.2\times 10^{-5}$ &
$(1.69\pm 0.15)\times 10^{-2}$ \\[1ex]
$\omega\eta$ & $5.7\times 10^{-3}$ & $7.0\times 10^{-6}$  &
$1.3\times 10^{-5}$ &
$(1.74\pm 0.20)\times 10^{-3}$ \\[1ex]
$\omega\eta^\prime$ & $4.2\times 10^{-3}$& $1.6\times 10^{-6}$ &
$1.5\times 10^{-5}$
& $(1.82\pm 0.21)\times 10^{-4}$ \\[1ex]
$\phi\eta$ &  $1.1\times 10^{-2}$ & $2.0\times 10^{-5}$ &
$2.0\times 10^{-5}$  &
$(7.4\pm 0.8)\times 10^{-4}$ \\[1ex]
$\phi\eta^\prime$ & $8.3\times 10^{-3}$ & $1.3\times 10^{-5}$ &
$1.8\times 10^{-5}$ &
$(4.0\pm 0.7)\times 10^{-4}$ \\[1ex]
$K^{*+}K^-+c.c.$ & $3.4\times 10^{-2}$ & $7.4\times 10^{-5}$ &
$5.9\times 10^{-5}$ &
$(5.0\pm 0.4)\times 10^{-3}$ \\[1ex]
$K^{*0}\bar{K^0}+c.c.$ & $7.8\times 10^{-2}$ & $1.6\times 10^{-4}$
& $1.3\times 10^{-4}$
& $(4.2\pm 0.4)\times 10^{-3}$ \\[1ex]\hline
\end{tabular}
\caption{ Branching ratios for $J/\psi\to\gamma^*\to VP$ without
the form factor (${\cal F}(q^2)=1$) and with the monopole (MP)
form factor. Column MP-C corresponds to process (a) and (b) in a
constructive phase with an effective mass $\Lambda=0.616$ GeV.
Column MP-D corresponds to (a) and (b) in a destructive phase with
$\Lambda=0.65$ GeV. The data for branching ratios are from
PDG2006~\cite{pdg2006}. } \label{tab-4}
\end{table}

\begin{table}[ht]
\begin{tabular}{c|c|c|c|c}
\hline Decay channels & ${\cal F}(q^2)=1$ & MP-D & MP-C  & Exp.
data  \\[1ex]\hline
$\rho\eta$ & $1.3\times 10^{-2}$ & $7.3\times 10^{-6}$ &
$1.5\times 10^{-5}$  &
$(2.2\pm 0.6)\times 10^{-5}$ \\[1ex]
$\rho\eta^\prime$ & $7.3\times 10^{-3}$ & $2.3\times 10^{-6}$ &
$1.1\times 10^{-5}$ &
$(1.9 \begin{array}{c} +1.7 \\ -1.2\end{array})\times 10^{-5}$ \\[1ex]
$\omega\pi$ & $3.1\times 10^{-2}$ & $3.2\times 10^{-5}$ &
$2.6\times 10^{-5}$ & $(2.1\pm 0.6)\times
10^{-5}$ \\[1ex]
$\phi\pi$ & $8.7\times 10^{-5}$ & $8.8\times 10^{-8}$ & $7.0\times
10^{-8}$
& $<4\times 10^{-6}$ \\[1ex]\hline
$\rho^0\pi^0$ & $3.8\times 10^{-3}$ & $3.9\times 10^{-6}$ &
$3.1\times 10^{-6}$ &
*** \\[1ex]
$\rho\pi$ & $9.6\times 10^{-3}$ & $9.8\times 10^{-6}$ & $7.9\times
10^{-6}$ &
$(3.2\pm 1.2)\times 10^{-5}$ \\[1ex]
$\omega\eta$ & $1.1\times 10^{-3}$ & $6.5\times 10^{-7}$ &
$1.3\times 10^{-6}$ &
$< 1.1\times 10^{-5}$ \\[1ex]
$\omega\eta^\prime$ & $8.9\times 10^{-4}$ & $4.0\times 10^{-7}$ &
$1.2\times 10^{-6}$
& $(3.2\begin{array}{c} +2.5\\ -2.1\end{array})\times 10^{-5}$ \\[1ex]
$\phi\eta$ & $2.2\times 10^{-3}$ & $1.9\times 10^{-6}$ &
$2.0\times 10^{-6}$ &
$(2.8\begin{array}{c} +1.0\\ -0.8\end{array})\times 10^{-5}$ \\[1ex]
$\phi\eta^\prime$ & $1.9\times 10^{-3}$ & $1.6\times 10^{-6}$ &
$1.8\times 10^{-6}$ &
$(3.1\pm 1.6)\times 10^{-5}$ \\[1ex]
$K^{*+}K^-+c.c.$ & $6.7\times 10^{-3}$ & $7.0\times 10^{-6}$ &
$5.6\times 10^{-6}$ &
$(1.7\begin{array}{c} +0.8\\ -0.7\end{array} )\times 10^{-5}$ \\[1ex]
$K^{*0}\bar{K^0}+c.c.$ & $1.5\times 10^{-2}$ & $1.6\times
10^{-5}$ & $1.3\times 10^{-5}$ & $(1.09\pm 0.20)\times 10^{-4}$ \\[1ex]\hline
\end{tabular}
\caption{ Branching ratios for $\psi^\prime\to\gamma^*\to VP$
without the form factor (${\cal F}(q^2)=1$) and with the monopole
(MP) form factor. The notations are the same as Table~\ref{tab-6}.
The stars ``***" in $\rho^0\pi^0$ channel denotes the
unavailability of the data.} \label{tab-5}
\end{table}

\begin{table}[ht]
\begin{tabular}{c|c|c|c|c}
\hline Decay channels & $R_1^{VP}(\%)$  & $R_2^{VP}(\%)$ &
$R_3^{VP}(\%)$ & Exp.
data (\%)  \\[1ex]\hline
$\rho\eta$ & 19 & 9 & 9 &
$11.5\pm 5.0 $\\[1ex]
$\rho\eta^\prime$ & 21  & 52 & 7 &
$23.5\pm 17.8$ \\[1ex]
$\omega\pi$ & 19  & 9 & 9 & $5.0\pm 1.8$\\[1ex]
$\phi\pi$ & 19  & 11 & 8 & $< 62.5 $ \\[1ex]\hline
$\rho^0\pi^0$ & 19  & 11 & 8 &
*** \\[1ex]
$\rho\pi$ & 19  & 11 & 8 &
$0.2\pm 0.1$ \\[1ex]
$\omega\eta$ & 19 & 10 & 8 &
$< 0.6\pm 0.1$ \\[1ex]
$\omega\eta^\prime$ & 21  & 25 & 8
& $18.5\pm 13.2$ \\[1ex]
$\phi\eta$ & 20 & 10 & 10 & $4.1\pm 1.6$ \\[1ex]
$\phi\eta^\prime$ & 23  & 13 & 10 &
$8.7\pm 5.5$ \\[1ex]
$K^{*+}K^-+c.c.$ & 20  & 9 & 9 &
$0.4\pm 0.2$ \\[1ex]
$K^{*0}K^0+c.c.$ & 20  & 9 & 9 &
$2.7\pm 0.7$ \\[1ex]\hline
\end{tabular}
\caption{  Branching ratio fractions for
$\psi^\prime\to\gamma^*\to VP$ over $J/\psi\to\gamma^*\to VP$.
$R_1^{VP}$ refers to ${\cal F}(q^2)=1$, while $R_2^{VP}$ and
$R_3^{VP}$ correspond to calculations with MP-D and MP-C form
factors, respectively. The last column is extracted from the
experimental date~\cite{pdg2006}. The stars ``***" in
$\rho^0\pi^0$ channel denotes the unavailability of the data. }
\label{tab-6}
\end{table}


\end{document}